\pdfoutput=1
\documentclass[journal]{IEEEtran}

\usepackage{cite}
\usepackage{amsmath,amssymb,amsfonts}
\usepackage{graphicx}
\usepackage{xcolor}
\usepackage{color}
\usepackage{algorithm}
\usepackage{algpseudocode}
\usepackage{multirow}
\usepackage{bm}
\usepackage[acronyms,shortcuts]{glossaries}
\usepackage{tabularx}
\usepackage{booktabs}
\usepackage{array}

\newcommand{\sumx}[2]{\sum\limits_{#1}^{#2}}
\newcommand{\bb}[1]{\mathbb{#1}}
\newcommand{\ten}[1]{\boldsymbol{\mathcal #1}}
\newcommand{\ma}[1]{\boldsymbol{#1}}
\newcommand{\diag}[1]{\text{diag}{#1}}
\newcommand{\nmode}[2]{[#1]_{(#2)}}
\newcommand{\fronorm}[1]{\left\|#1\right\|_{\text{F}}}
\renewcommand{\vec}[1]{\text{vec}{(#1)}}

\newacronym{2G}{2G}{second generation}
\newacronym{3G}{3G}{third generation}
\newacronym{4G}{4G}{fourth generation}
\newacronym{5G}{5G}{fifth generation}
\newacronym{B5G}{B5G}{beyond fifth generation}
\newacronym{6G}{6G}{sixth generation}
\newacronym{3GPP}{3GPP}{3$\text{rd}$~Generation Partnership Project}
\newacronym{LTE}{LTE}{long term evolution}
\newacronym{NR}{NR}{new radio}
\newacronym{LS}{LS}{least squares}

\newacronym{IRS}{IRS}{intelligent reconfigurable surface}
\newacronym{RIS}{RIS}{reconfigurable intelligent surface}
\newacronym{LIS}{LIS}{large intelligent surface}
\newacronym{SDS}{SDS}{software-defined surface}

\newacronym{D2D}{D2D}{device-to-device}
\newacronym{BS}{BS}{base station}
\newacronym{UE}{UE}{user equipment}

\newacronym{SU}{SU}{single-user}
\newacronym{MU}{MU}{multi-user}
\newacronym{SISO}{SISO}{single-input single-output}
\newacronym{MISO}{MISO}{multiple-input single-output}
\newacronym{SIMO}{SIMO}{single-input multiple-output}
\newacronym{MIMO}{MIMO}{multiple-input multiple-output}

\newacronym{CSI}{CSI}{channel state information}
\newacronym{LOS}{LOS}{line-of-sight}
\newacronym{NLOS}{NLOS}{non-line-of-sight}

\newacronym{QoS}{QoS}{quality-of-service}
\newacronym{SE}{SE}{spectral efficiency}
\newacronym{EE}{EE}{energy efficiency}
\newacronym{SINR}{SINR}{signal to interference plus noise ratio}
\newacronym{SNR}{SNR}{signal to noise ratio}

\newacronym{ProSe}{ProSe}{proximity services}
\newacronym{NSPS}{NSPS}{national security and public safety}

\newacronym{RRM}{RRM}{radio resource management}
\newacronym{MS}{MS}{mode selection}
\newacronym{RA}{RA}{resource allocation}
\newacronym{PC}{PC}{power control}

\newacronym{BCD}{BCD}{block coordinate descent}

\newacronym{RF}{RF}{radio frequency}
\newacronym{AWGN}{AWGN}{additive white Gaussian noise}
\newacronym{MRC}{MRC}{maximum ratio combining}

\newacronym{AF}{AF}{amplify-and-forward}
\newacronym{DF}{DF}{decode-and-forward}
\newacronym{DFT}{DFT}{discrete Fourier transform}
\newacronym{Tx}{Tx}{transmitter}
\newacronym{Rx}{Rx}{receiver}
\newacronym{ALS}{ALS}{alternating least squares}
\newacronym{SVD}{SVD}{singular value decomposition}
\newacronym{HOSVD}{HOSVD}{higher order singular value decomposition}
\newacronym{T-HOSVD}{T-HOSVD}{truncated higher-order singular value decomposition}
\newacronym{PARAFAC}{PARAFAC}{PARAllel FACtors}
\newacronym{AOD}{AOD}{angle of departure}
\newacronym{AOA}{AOA}{angle of arrival}
\newacronym{URA}{URA}{uniform rectangular array} 
\newacronym{ADR}{ADR}{achievable data rate}
\newacronym{NMSE}{NMSE}{normalized mean square error}
\newacronym{SER}{SER}{symbol error rate}
\newacronym{LRA}{LRA}{low-rank approximation}
\newacronym{JD}{JD}{joint detection}

\newacronym{ULA}{ULA}{uniform linear array}
\newacronym{mmWave}{mmWave}{milimiter-wave}
\newacronym{CS}{CS}{compressed sensing}
\newacronym{OFDM}{OFDM}{orthogonal frequency division multiplexing}
\newacronym{PIN}{PIN}{positive-intrinsic-negative}
\newacronym{FLOPS}{FLOPS}{floating-point operations per second}
\newacronym{TAO}{TAO}{tensor alternating optimization}
\newacronym{MS-TAO}{MS-TAO}{multi-stream tensor alternating optimization}
\newacronym{AO}{AO}{ alternating optimization}
\newacronym{PGM}{PGM}{ projected gradient method}

\title{Low-Complexity Tensor Beamforming for RIS-Aided Multiuser Multistream MIMO Systems}

\author{Bruno Sokal,~\IEEEmembership{Member,~IEEE,} André L.F. de Almeida,~\IEEEmembership{Senior Member,~IEEE,} and Martin Haardt,~\IEEEmembership{Fellow,~IEEE}%
\thanks{Bruno Sokal and Andr\'{e} L.  F. de Almeida are with the Wireless Telecom Research Group (GTEL), Department of Teleinformatics Engineering, Federal University of Cear\'{a}, Fortaleza-CE, Brazil. E-mails: \{brunosokal,andre\}@gtel.ufc.br. Martin Haardt is with the Communications Research Laboratory, Ilmenau University of Technology, Ilmenau, Germany. E-mail: martin.haardt@tu-ilmenau.de. This work is partially supported by the National Institute of Science
and Technology (INCT-Signals) sponsored by Brazil’s National Council for Scientific and Technological Development (CNPq) (Proc. 406517/2022-3), FUNCAP (Proc. ITR-0214-00041.01.00/23), and partially supported by CAPES/PROBRAL under grant  88881.894956/2023-01. The authors gratefully acknowledge the partial support of the German Research Foundation (DFG) under Grant 402834619 (AdAMMM-II, HA 2239/14-3). This work has been submitted to the IEEE for possible publication. Copyright may be transferred without notice, after which this version may no longer be accessible.
}
}

\begin{document}

\maketitle

\begin{abstract}

We address joint active and passive beamforming for uplink RIS-assisted multi-user multi-stream MIMO systems with joint detection. The coupled design of the receive combiner, block-diagonal user precoders, and RIS phase vector is formulated through a third-order composite channel tensor. Exploiting this multilinear structure, we propose a multi-stream tensor alternating optimization method that updates the combiner, user precoders, and
RIS coefficients via low-dimensional tensor projections. Simulations show that the proposed method approaches a multi-start
alternating-optimization benchmark while reducing computational complexity and improving large-RIS scaling.
\end{abstract}


\section{Introduction}
Reconfigurable intelligent surfaces (RISs) have emerged as a promising technology for improving wireless links by shaping the propagation environment through a large number of nearly passive reflecting elements with adjustable phase responses \cite{wu2019intelligent,huang2019reconfigurable,bjornson2020reconfigurable}. By properly configuring the RIS phase shifts, the end-to-end channel can be adapted to favor reliable spatial multiplexing and enhance the spectral efficiency of multiple-input multiple-output (MIMO) systems. In uplink multi-user MIMO scenarios, this capability is particularly attractive because the receiver can jointly process the signals transmitted by several users. When each user transmits multiple data streams, however, the design problem becomes substantially more challenging: the users' precoders, the common RIS phase vector, and the receive combiner used as a linear front-end for \ac{JD} must be optimized in a coupled manner under non-convex unit-modulus constraints.

Several works have addressed active and passive beamforming for RIS-assisted wireless systems. The seminal formulation in \cite{wu2019intelligent} demonstrated the effectiveness of alternating optimization for joint transmit and RIS beamforming, while \cite{huang2019reconfigurable} studied energy-efficient RIS-aided communication. For multi-user RIS-assisted systems, weighted sum-rate and fractional-programming-based designs have been proposed in \cite{guo2020weighted,MA2021}, providing important benchmarks for multi-user beamforming with RISs. In the uplink multi-user MIMO case, \cite{you2020energy} investigated the spectral-efficiency and energy-efficiency tradeoff under joint active and passive beamforming. Lower-complexity RIS beamforming strategies have also been developed, for instance through codebook- and correlation-based designs \cite{wu2022lowcomplexity}. Nevertheless, these methods are mainly formulated in the matrix domain and do not explicitly exploit the multilinear structure of the composite RIS channel.

Tensor-based signal processing provides a natural framework for RIS-assisted MIMO systems because the received signals and cascaded channels are inherently indexed by multiple domains, such as the transmit array, receive array, users, pilots, and RIS elements. Tensor formulations have been successfully used for channel estimation in RIS-assisted MIMO and multi-user MIMO systems, enabling structured representations and factor decoupling of the involved channels \cite{Gil2021,WeiLi:2021,ardah2021trice,gil2022}. However, after channel estimation, the beamforming stage is often performed by matricizing the estimated channels, which may discard useful multidimensional structure. Recently, tensor-based beamforming was shown to be effective for single-user multi-stream RIS-assisted MIMO systems \cite{sokal2025tensorRIS}. Extending this idea to the uplink multi-user multi-stream case is nontrivial because the transmit-side beamformer must preserve a block-diagonal structure across users, whereas the RIS phase vector and the JD receive combiner are common to all users and streams.

Motivated by this gap, this letter develops a low-complexity tensor-based beamforming method for uplink RIS-assisted multi-user multi-stream MIMO systems with JD at the receiver. We reshape the composite RIS channel into a third-order tensor whose modes correspond to the receive array, the aggregate users' transmit array, and the RIS domain, and use this representation to derive a multilinear surrogate for the joint design of the JD combiner, the users' block-diagonal precoders, and the RIS phase vector. Based on this formulation, we propose the \ac{MS-TAO} algorithm, which updates the receiver combiner, the per-user precoders, and the RIS coefficients through low-dimensional tensor projections while preserving the multi-user precoder constraints of the uplink model.

Numerical results show that MS-TAO achieves a favorable spectral-efficiency/complexity tradeoff, approaching the performance of multi-start alternating-optimization benchmark with a substantially lower computational complexity and more favorable scaling with the number of RIS elements. These results indicate that tensor-based beamforming is a suitable tool for scalable joint active and passive beamforming in RIS-assisted uplink multi-user multi-stream MIMO systems.

\section{System Overview}
\label{Sec:system Model}

We consider an uplink multi-user \ac{MIMO} \ac{RIS}-assisted communication scenario with $K$ \acp{UE}, where the $k$th UE is equipped with $M_{T,k}$ antennas and transmits $R_k$ data streams. The \ac{Rx} has $M_R$ antennas and the RIS comprises $N$ passive reflecting elements. We define the aggregate transmit dimension and the total number of data streams as $M_T=\sum_{k=1}^{K}M_{T,k}$ and $R=\sum_{k=1}^{K}R_k$, respectively. In this setup, the direct communication link between the \acp{UE} and the \ac{Rx} is assumed to be blocked or negligible. To estimate the cascaded channels (Tx-RIS and RIS-Rx), the UEs transmit orthogonal pilot sequences, and the RIS applies a predefined space-time reflection pattern across its elements. Assuming a two-time-scale protocol \cite{gil2022}, in which the pilot sequences are transmitted over $I$ blocks, each with a duration of $T$ time slots, the received signal in the $t$th time slot of the $i$th block is given by
\begin{align}\label{eq:rec_sig_t}
\ma y_{i,t} &= \ma G \diag (\ma s_i)\sumx{k=1}{K}\ma H_k \ma x_{k,t} + \ma b_{i,t}, \\
&= \ma{G}\diag (\ma s_i)\ma{H}\ma{x}_t  + \ma b_{i,t},
\end{align}
where $\ma{x}_{k,t}\in\mathbb{C}^{M_{T,k}\times1}$ and $\ma{x}_t = [\ma x_{1,t}^{\text{T}},\ldots,\ma{x}_{K,t}^{\text{T}}]^{\text{T}} \in \mathbb{C}^{M_T \times 1}$ denotes the pilot vector transmitted by all $K$ UEs in the $t$th time slot. The vector $\ma s_i = \left[a_{1,i}e^{j\phi_{1,i}},\dots,a_{N,i}e^{j\phi_{N,i}}\right]^{\textrm{T}} \in \mathbb{C}^{N \times 1}$ denotes the RIS reflection coefficients applied during the $i$th block, where $\phi_{n,i} \in (0,2\pi]$ is the phase shift introduced by the $n$th RIS element and $a_{n,i}=1$ for all $n$ and $i$. The matrix $\ma G \in \mathbb{C}^{M_R \times N}$ denotes the channel between the \ac{Rx} and the RIS, while $\ma{H} = [\ma H_{1},\ldots, \ma{H}_K] \in \mathbb{C}^{N \times M_T}$ stacks the UE--RIS channels $\ma H_k\in\mathbb{C}^{N\times M_{T,k}}$. The vector $\ma b_{i,t} \in \mathbb{C}^{M_R \times 1}$ models the \ac{AWGN} at the receiver. We assume a block-fading scenario, meaning that the channel matrices $\ma H$ and $\ma G$ remain constant during the $TI$ time slots. Next, we describe a channel estimation procedure that enables channel separation, which is useful for beamformer optimization during the data-transmission phase.

\subsection{Channel estimation}
Let us assume that the UEs transmit orthogonal pilots, i.e., $\ma{X}\ma{X}^{\text{H}} = \ma{I}_{M_T}$, with $\ma{X} = [\ma{x}_1,\ldots,\ma{x}_T] \in \bb{C}^{M_T \times T}$ and $T\geq M_T$, and that, during channel estimation, the RIS configuration for each block $i$ is chosen from a predefined reflection-pattern matrix. Collecting the $T$ time slots, the received signal $\ma{Y}_{i} \in \mathbb{C}^{M_R\times T}$ at the $i$th block is given by
\begin{align}
   \ma{Y}_{i} &=[\ma y_{i,1},\ldots,\ma y_{i,T}]  = \ma{G}\diag (\ma s_i)\ma{H}\ma{X} + \ma{B}_i,
\end{align}
with $\ma{B}_i = [\ma b_{i,1},\ldots,\ma{b}_{i,T}]$. Using a matched filter to remove the pilot signal, the filtered received signal is expressed as
\begin{align}
   \bar{\ma{Y}}_{i} &= \ma{Y}_i\ma{X}^{\text{H}} = \ma{G}\diag (\ma s_i)\ma{H} + \bar{\ma{B}}_i,
\end{align}
with $\bar{\ma{B}}_i = \ma{B}_i\ma{X}^{\text{H}}$ being the filtered \ac{AWGN}. As in \cite{Gil2021}, the filtered received signal can be interpreted as a frontal slice of a PARAFAC tensor, given by
\begin{align}
  \label{eq:tenY}  \bar{\ten{Y}} = \ten{I}_{3,N} \times_1 \ma{G} \times_2 \ma{H}^{\text{T}} \times_3 \ma{S} \, \, + \bar{\ten{B}} \in \bb{C}^{M_R \times M_T \times I},
\end{align}
where $\ma{S}\in\bb{C}^{I\times N}$ collects the RIS training patterns with $[\ma{S}]_{i,:}=\ma{s}_i^{\text{T}}$. The tensor $\ten{I}_{3,N} \in \bb{R}^{N \times N \times N}$ is a superdiagonal tensor whose entries $\ten{I}_{3,N}[n^{\prime},n^{\prime\prime},n^{\prime\prime\prime}] = 1$ when $n^{\prime} = n^{\prime\prime} = n^{\prime\prime\prime}$ and zero otherwise, for $\{n^{\prime},n^{\prime\prime},n^{\prime\prime\prime}\} = 1,\ldots, N$. As explained in \cite{Gil2021}, a low-overhead iterative algorithm that exploits the tensor model in \eqref{eq:tenY} can be derived to obtain decoupled estimates of the channels $\ma{G}$ and $\ma{H}$. 

\subsection{Data phase transmission}
After estimating the channels, the \ac{Rx} optimizes the beamformers for the data-transmission phase. Assuming a linear receive front-end followed by \ac{JD} over the $R$ streams, the post-combining signal can be expressed as
\begin{align}
    \ma{y} = \sqrt{\frac{P}{R}}\ma{W}^{\text{H}}\ma{G}\diag{(\ma{s})}\ma{H}\ma{Q}\ma{x} + \ma{W}^{\text{H}}\ma{b},
\end{align}
where $\ma{W} \in \bb{C}^{M_R \times R}$ is the combiner at the \ac{Rx}, $\ma{s} \in \bb{C}^{N \times 1}$ is the optimized beamforming vector for the RIS, and $\ma{Q} = \text{blkdiag}(\ma{Q}_1,\ldots,\ma{Q}_K) \in \bb{C}^{M_T \times R}$, with $\ma{Q}_k \in \bb{C}^{M_{T,k} \times R_k}$ denoting the precoder of the $k$th user. Assuming Gaussian signaling, i.e., $\mathbb{E}[\ma{x}\ma{x}^{\rm H}]=\ma{I}_{R}$, an orthogonal combiner $\ma{W}$, i.e., $\ma{W}^{\text{H}}\ma{W}=\ma{I}_{R}$, and defining $\rho=P/\sigma^2$, the achievable sum rate with \ac{JD} at the output is \cite{telatar1999capacity,tse2005fundamentals}
\begin{align}
\label{eq:rate_JD}
R_{\rm sum}
=
\log_2\det\!\Big(
\ma{I}_{R}
+
\frac{\rho}{R}\ma{W}^{\text{H}}
\ma{H}_{\rm eq}\ma{Q}\ma{Q}^{\text{H}}\ma{H}_{\rm eq}^{\text{H}}\ma{W}\Big). 
\end{align}
where $\ma{H}_{\rm eq} = \ma{G}\diag{(\ma{s})}\ma{H}$ and $\ma{W}^{\text{H}}\mathbb{E}[\ma{b}\ma{b}^{\rm H}]\ma{W}=\sigma^2 \ma{I}_{R}$. 

In this work, we aim to maximize the achievable sum rate in \eqref{eq:rate_JD}. However, the log-det objective leads to a highly non-convex optimization problem with respect to the RIS coefficients $\ma{s}$ and the beamforming matrices $\ma{W}$ and $\ma{Q}$. To overcome this difficulty, we instead approximate the problem by the following surrogate:
\begin{align}\label{eq:problem_frob}
&\underset{\ma{Q},\ma{W},\ma{s}}{\text{max}} 
 \fronorm{\ma{W}^{\text{H}}\ma{G}\diag(\ma{s})\ma{H}\ma{Q}}^2, \, \, \,\text{s.t.} \,\,
 \ma{W}^{\text{H}}\ma{W} = \ma{I}_{R},\\
\notag& \ma{Q}_k^{\text{H}}\ma{Q}_k = \ma{I}_{R_k}, \,\,k=1,\ldots,K, \,\,\, |s_n| = 1, \,\, n=1,\ldots,N .
\end{align}
In particular, maximizing the Frobenius norm of the effective channel provides a meaningful surrogate for maximizing the achievable sum rate in the low-SNR regime~\cite{telatar1999capacity,tse2005fundamentals,verdu2002spectral}.

Similar surrogate formulations based on maximizing the received signal power or the Frobenius norm of the effective channel have also been widely adopted in the literature to simplify beamforming and RIS optimization problems, particularly when the original log-det capacity expression leads to intractable non-convex design problems \cite{wu2019intelligent,bjornson2020reconfigurable,huang2019reconfigurable}. Next, we propose a tensor-based method that jointly designs the system's beamformers.
\section{Tensor-Based Problem Formulation}
After the channel estimation phase, the \ac{Rx} optimizes the system beamformers for data transmission. For simplicity, we focus on the beamforming design assuming that the combined channel  $\ma{T} = \ma{H}^{\text{T}} \diamond \ma{G} \in \bb{C}^{M_RM_T \times N}$ is available at the receiver. Using the property $\text{vec}(\ma{ABC}) = (\ma{C}^{\text{T}} \otimes \ma{A})\text{vec}(\ma{B})$, we have
\begin{align}
\vec{\ma{W}^{\text{H}}\ma{G}\diag(\ma{s})\ma{H}\ma{Q}} = (\ma{Q}^{\text{T}} \otimes \ma{W}^{\text{H}})\ma{T}\ma{s},
\end{align}
the problem in \eqref{eq:problem_frob} can be rewritten as
\begin{align}\label{eq:problem_ten_frob}
&\underset{\ma{Q},\ma{W},\ma{s}}{\text{max}} 
 \fronorm{\ten{T} \times_1 \ma{W}^{\text{H}}\times_2 \ma{Q}^{\text{T}} \times_3 \ma{s}^{\text{T}} }^2, \, \, \,\text{s.t.} \,\,
 \ma{W}^{\text{H}}\ma{W} = \ma{I}_{R},\\
\notag& \ma{Q}_k^{\text{H}}\ma{Q}_k = \ma{I}_{R_k}, \,\,k=1,\ldots,K, \,\,\, |s_n| = 1, \,\, n=1,\ldots,N ,
\end{align}
where $\ten{T} \in \bb{C}^{M_R \times M_T \times N}$ is the tensorized form of the combined channel $\ma{T} \in \bb{C}^{M_RM_T \times N}$. More specifically, we map the elements of the combined channel as $\ten{T}_{[i],[j],[n]} = \ma{T}_{[ i +(j-1)M_R],[n] }$, with $i = 1,\ldots,M_R$, $j = 1,\ldots,M_T$, and $n = 1,\ldots,N$. By recasting \eqref{eq:problem_frob} as \eqref{eq:problem_ten_frob}, we explicitly exploit the tensor structure of the combined channel by separating the receiver dimension ($M_R$), the aggregate transmit dimension ($M_T$), and the RIS spatial dimension ($N$). This representation enables structured joint optimization of the beamformers.

\begin{table*}[!t]
\centering
\refstepcounter{table}\label{tab:complexity_tao_baseline}
\centerline{\footnotesize TABLE~\thetable: DOMINANT COMPUTATIONAL COMPLEXITY}
\vspace*{0.05cm}
\footnotesize
\renewcommand{\arraystretch}{1.25}
\setlength{\tabcolsep}{3pt}
\begin{tabular}{
p{0.24\textwidth}
p{0.28\textwidth}
p{0.22\textwidth}
p{0.18\textwidth}
}
\hline
\textbf{Algorithm} &
\textbf{Dominant complexity} &
\textbf{Required CSI} &
\textbf{Scaling in $N$} \\
\hline

Proposed MS-TAO &
$
\mathcal{O}\!\left(
I_{\rm TAO} \, N R^4
\right)
$
&
Composite channel tensor $\ten{T}$ &
Linear \\



\hline

Multi-start AO benchmark &
$
\mathcal{O}\!\left(
N_{\rm st} I_{\rm BL} \, N^2 (R+I_{\rm RIS})
\right)
$
&
Separated channels $\mathbf{G}$ and $\mathbf{H}$ &
Quadratic \\

\hline

Wu--Liu-inspired codebook \cite{wu2022lowcomplexity} &
$
\mathcal{O}\!\left(
N_{\rm c} I_{\rm WQ} \, N^2 R
\right)
$
&
Separated channels $\mathbf{G}$ and $\mathbf{H}$ &
$N_{\rm c}N^2$ ($N^3$ if $N_{\rm c}=N$) \\
\hline
\end{tabular}
\end{table*}

\begin{algorithm}[!t]
\caption{Proposed \ac{MS-TAO}}
\label{alg:tao_msmt}
\begin{algorithmic}[1]
    \State \textbf{Input:} Combined channel tensor $\ten{T}=[\ten{T}_1 \sqcup_2 \cdots \sqcup_2 \ten{T}_K] \in \bb{C}^{M_R \times M_T \times N}$, stream dimensions $\{R_k\}_{k=1}^{K}$ with $R=\sum_{k=1}^{K}R_k$, maximum number of iterations $I_{\max}$, and tolerance $\epsilon$
    \For{$k=1,\ldots,K$}
        \State Compute the \ac{SVD} of $\nmode{\ten{T}_k}{2}$ and set $\ma{Q}_k^{(0)} = \left([\ma{U}_{Q,k}^{(0)}]_{:,1:R_k}\right)^*$
    \EndFor
    \State Form $\ma{Q}^{(0)} = \mathrm{blkdiag}(\ma{Q}_1^{(0)},\ldots,\ma{Q}_K^{(0)})$
    \State Compute the \ac{SVD} of $\nmode{\ten{T}}{3}$ and set $\ma{s}^{(0)} = e^{-j\angle [\ma{U}_{s}^{(0)}]_{:,1}}$
    \For{$i=1,\ldots,I_{\max}$}
        \State Compute $\ten{T}_{W}^{(i)} = \ten{T} \times_2 \left(\ma{Q}^{(i-1)}\right)^{\rm T} \times_3 \left(\ma{s}^{(i-1)}\right)^{\rm T}$
        \State Compute the \ac{SVD} of $\nmode{\ten{T}_{W}^{(i)}}{1}$ and set $\ma{W}^{(i)} = [\ma{U}_{W}^{(i)}]_{:,1:R}$
        \For{$k=1,\ldots,K$}
            \State Compute $\ten{T}_{Q,k}^{(i)} = \ten{T}_k \times_1 \left(\ma{W}^{(i)}\right)^{\rm H} \times_3 \left(\ma{s}^{(i-1)}\right)^{\rm T}$
            \State Compute the \ac{SVD} of $\nmode{\ten{T}_{Q,k}^{(i)}}{2}$ and set $\ma{Q}_k^{(i)} = \left([\ma{U}_{Q,k}^{(i)}]_{:,1:R_k}\right)^*$
        \EndFor
        \State Form $\ma{Q}^{(i)} = \mathrm{blkdiag}(\ma{Q}_1^{(i)},\ldots,\ma{Q}_K^{(i)})$
        \State Compute $\ten{T}_{s}^{(i)} = \ten{T} \times_1 \left(\ma{W}^{(i)}\right)^{\rm H} \times_2 \left(\ma{Q}^{(i)}\right)^{\rm T}$
        \State Compute the \ac{SVD} of $\nmode{\ten{T}_{s}^{(i)}}{3}$ and set $\ma{s}^{(i)} = e^{-j\angle [\ma{U}_{s}^{(i)}]_{:,1}}$
        \State Evaluate the stopping criterion $f^{(i)}$ using \eqref{eq:stop_crit}
        \If{$i>1$ and $\left|f^{(i)}-f^{(i-1)}\right| / \left| f^{(i-1)}\right| < \epsilon$}
            \State \textbf{break}
        \EndIf
    \EndFor
    \State \textbf{return} $\ma{W}^{(i)}$, $\ma{Q}^{(i)}$, and $\ma{s}^{(i)} $.
\end{algorithmic}
\end{algorithm}

\section{Proposed Tensor Alternating Method}
In this section, we propose a tensor alternating optimization method for the joint design of the receive combiner, the users' precoders, and the RIS phase vector for the multi-user multi-stream uplink scenario. The proposed method exploits the multilinear structure of the combined channel tensor $\ten{T} \in \bb{C}^{M_R \times M_T \times N}$ and seeks to improve the surrogate objective in \eqref{eq:problem_ten_frob}. In contrast to the single-user multi-stream case, the multi-user structure imposes a block-diagonal constraint on the global precoder $\ma{Q}=\mathrm{blkdiag}(\ma{Q}_1,\ldots,\ma{Q}_K)$, so the combiner $\ma{W}$ and the RIS vector $\ma{s}$ are updated globally, whereas each precoder $\ma{Q}_k$ is updated separately.

Assume that, at the $i$th iteration, the previous estimates $\ma{Q}^{(i-1)}$ and $\ma{s}^{(i-1)}$ are available. The combiner update is obtained by defining
\begin{align*}
    \ten{T}_{W}^{(i)} = \ten{T} \times_2 \left(\ma{Q}^{(i-1)}\right)^{\rm T} \times_3 \left(\ma{s}^{(i-1)}\right)^{\rm T} \in \bb{C}^{M_R \times R \times 1}.
\end{align*}
Then, from the \ac{SVD} of the mode-$1$ unfolding,  $\nmode{\ten{T}_{W}^{(i)}}{1} = \ma{U}_{W}^{(i)}\ma{\Sigma}_{W}^{(i)}\ma{V}_{W}^{(i)\rm H}$, we define the receiver combiner as the dominant left singular subspace,
\begin{align}
    \ma{W}^{(i)} = [\ma{U}_{W}^{(i)}]_{:,1:R} \in \bb{C}^{M_R \times R}.
\end{align}

For a fixed $\ma{W}^{(i)}$ and $\ma{s}^{(i-1)}$, each user's precoder is updated independently. Let $\ten{T}_k \in \bb{C}^{M_R \times M_{T,k} \times N}$ denote the subtensor associated with the $k$th user. For  $k=1,\ldots,K$, define
\begin{align*}
    \ten{T}_{Q,k}^{(i)} = \ten{T}_k \times_1 \left(\ma{W}^{(i)}\right)^{\rm H} \times_3 \left(\ma{s}^{(i-1)}\right)^{\rm T} \in \bb{C}^{R \times M_{T,k} \times 1}.
\end{align*}
Using the \ac{SVD} of its mode-$2$ unfolding, $\nmode{\ten{T}_{Q,k}^{(i)}}{2} = \ma{U}_{Q,k}^{(i)}\ma{\Sigma}_{Q,k}^{(i)}\ma{V}_{Q,k}^{(i)\rm H}$,
the precoder of the $k$th user is given by
\begin{align}
    \ma{Q}_k^{(i)} = \left([\ma{U}_{Q,k}^{(i)}]_{:,1:R_k}\right)^* \in \bb{C}^{M_{T,k} \times R_k}.
\end{align}
After all user blocks are updated, the global precoder is reconstructed as $\ma{Q}^{(i)} = \mathrm{blkdiag}(\ma{Q}_1^{(i)},\ldots,\ma{Q}_K^{(i)})$. Finally, for fixed $\ma{W}^{(i)}$ and $\ma{Q}^{(i)}$, the RIS vector is updated from
\begin{align}
    \ten{T}_{s}^{(i)} = \ten{T} \times_1 \left(\ma{W}^{(i)}\right)^{\rm H} \times_2 \left(\ma{Q}^{(i)}\right)^{\rm T} \in \bb{C}^{R \times R \times N}.
\end{align}
From $\nmode{\ten{T}_{s}^{(i)}}{3} = \ma{U}_{s}^{(i)}\ma{\Sigma}_{s}^{(i)}\ma{V}_{s}^{(i)\rm H}$, the RIS vector is updated by phase projection: $\tilde{\ma{s}}^{(i)} = [\ma{U}_{s}^{(i)}]_{:,1},$ with $ \ma{s}^{(i)} = e^{-j\angle \tilde{\ma{s}}^{(i)}} $.

The three steps are repeated until convergence. A practical stopping criterion is based on the relative change of the surrogate objective,
\begin{align}
  \label{eq:stop_crit}  f^{(i)} = \fronorm{\ten{T} \times_1 \left(\ma{W}^{(i)}\right)^{\rm H} \times_2 \left(\ma{Q}^{(i)}\right)^{\rm T} \times_3 \left(\ma{s}^{(i)}\right)^{\rm T}}^2,
\end{align}
and the iterations are stopped when $i>1$ and the relative change between two consecutive values becomes smaller than a prescribed threshold $\epsilon$. The resulting procedure is summarized in Algorithm~\ref{alg:tao_msmt}.

\section{Computational Complexity Analysis}

Table~\ref{tab:complexity_tao_baseline} summarizes the dominant computational complexity of the considered methods.
For the proposed MS-TAO method, each iteration updates the receive combiner, the block-diagonal transmit precoder, and the RIS phase vector through tensor projections. For fixed $M_R$, $M_T$, $K$, and $R$, the large-RIS scaling is dominated by the RIS-mode update, where an $N\times R^2$ unfolding is processed. Since $N\gg R^2$, the corresponding economy-size SVD scales as $\mathcal{O}(NR^4)$. 
Hence, the overall complexity of MS-TAO is $\mathcal{O}(I_{\rm TAO}NR^4)$.

The multi-start AO benchmark repeats  for $N_{\rm st}$ independent random initializations of the RIS phase vector and retains the solution with the largest final objective value. It operates on the separated channels $\mathbf{G}$ and $\mathbf{H}$. Although the receive combiner and user precoders are updated using standard SVD/eigendecomposition steps, the dominant cost comes from the RIS phase update. In particular, the RIS update requires forming and repeatedly applying an $N\times N$ quadratic matrix associated with the unit-modulus phase optimization. The construction of this RIS-side matrix scales as $\mathcal{O}(N^2R)$, while the $I_{\rm RIS}$ inner phase-projection iterations scale as $\mathcal{O}(I_{\rm RIS}N^2)$. Therefore, the dominant complexity of the Frobenius-AO baseline is $\mathcal{O}\!\left(N_{\rm st}I_{\rm BL}N^2(R+I_{\rm RIS}) \right)$.

The original Wu--Liu codebook \cite{wu2022lowcomplexity} method reports a complexity of $\mathcal{O}(N_{\rm c})$ for the unified design  where \(N_{\rm c}\)
is the number of RIS codewords. In our adapted implementation, for each of the \(N_c\) RIS codewords, the active beamformers are refined for \(I_{\rm WQ}\)
iterations. This refinement uses RIS-domain quadratic forms, whose dominant cost scales as \(\mathcal{O}(N^2R)\) per
 codeword and active-beamforming iteration. Therefore, the implemented Wu--Liu-inspired baseline has dominant cost
\(\mathcal{O}(N_c I_{\rm WQ}N^2R)\).

Overall, the proposed MS-TAO method has the most favorable large-RIS scaling among the considered iterative methods, since it avoids the $N\times N$ RIS-side quadratic updates by exploiting the intrinsic tensor structure of the combined channel. 
\vspace{-0.55cm}
\section{Numerical Results}\label{Sec:Simulation_Results}

\begin{figure*}[!t]
\centering
\begin{minipage}[t]{0.32\textwidth}
    \centering
    \includegraphics[width=\linewidth]{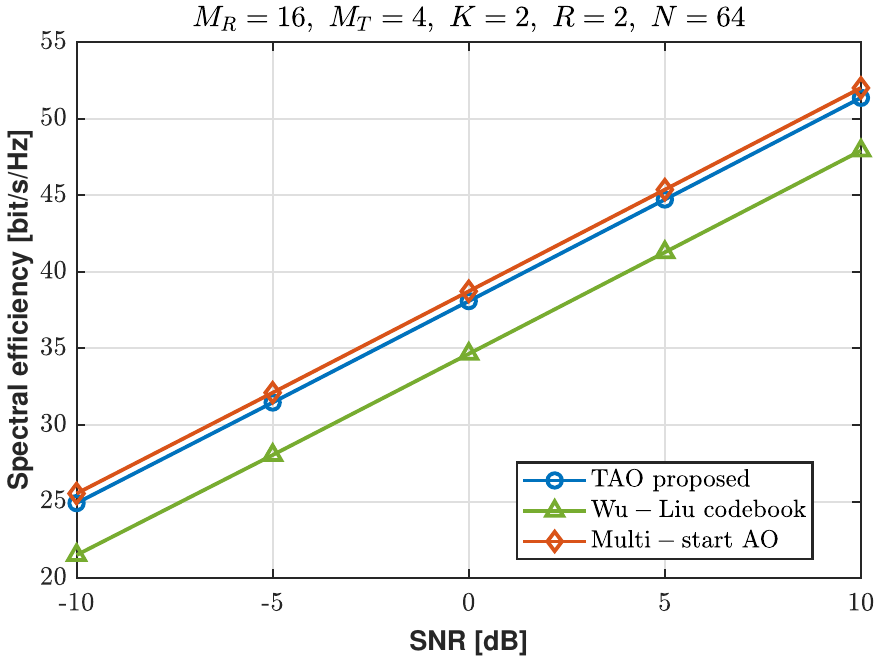}
    \vspace{-1mm}
    \centerline{(a)}
\end{minipage}
\hfill
\begin{minipage}[t]{0.32\textwidth}
    \centering
    \includegraphics[width=\linewidth]{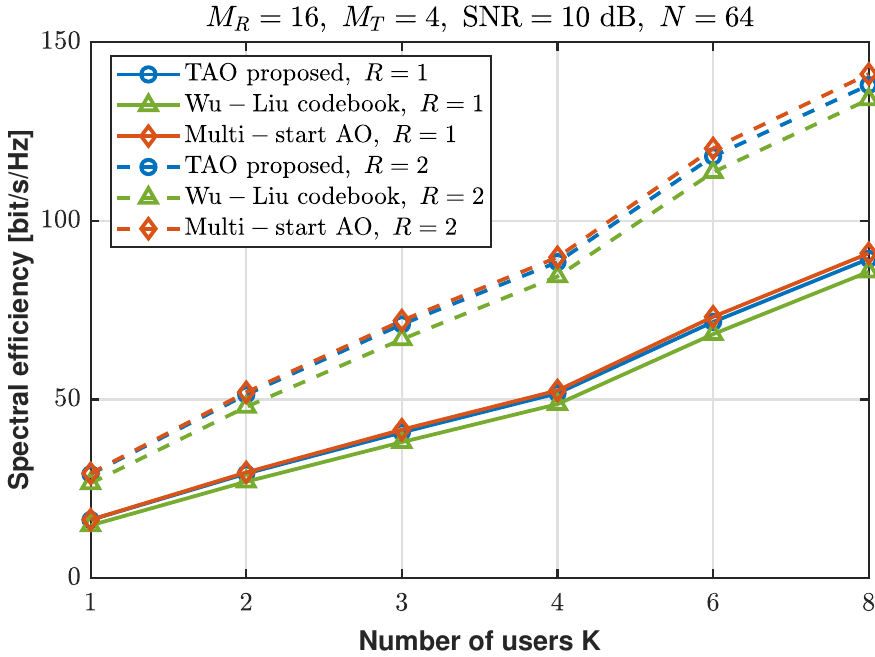}
    \vspace{-1mm}
    \centerline{(b) }
\end{minipage}
\hfill
\begin{minipage}[t]{0.32\textwidth}
    \centering
    \includegraphics[width=\linewidth]{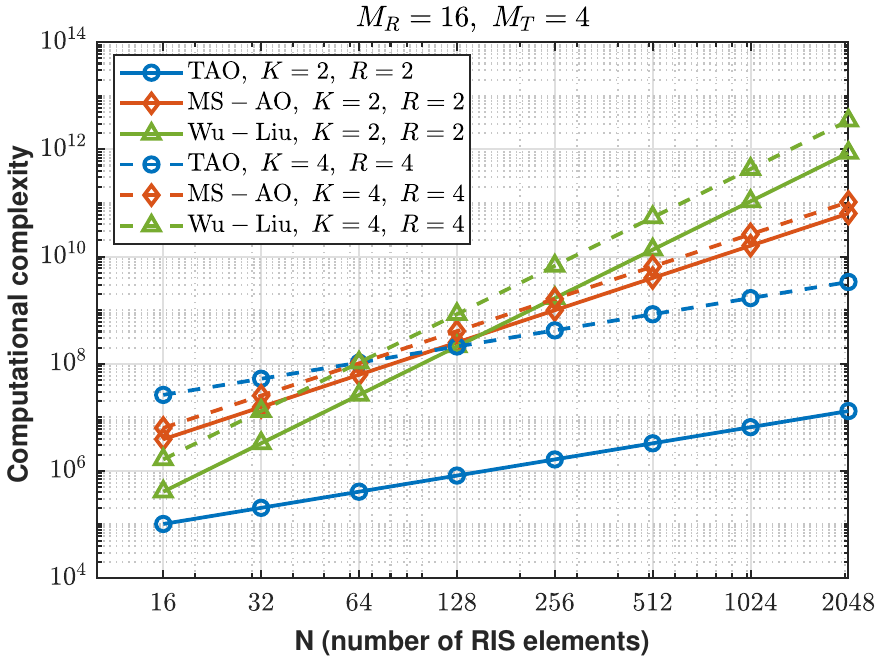}
    \vspace{-1mm}
    \centerline{(c) }
\end{minipage}

\caption{Comparison of the proposed MS-TAO method with benchmarks. 
(a) Sum spectral efficiency versus SNR for fixed $K$ and $R_{\rm UE}$. 
(b) Sum spectral efficiency versus the number of users $K$ for fixed SNR and different values of $R_{\rm UE}$. 
(c) Dominant computational complexity versus the number of RIS elements $N$.}
\label{fig:main_results}
\end{figure*}

Unless otherwise stated, we consider the equal-size special case $M_{T,k}=4$ and $R_k=R_{\rm UE}$ for all $k$, so that $M_T=4K$ and $R=KR_{\rm UE}$ in the general model. We use $M_R=16$ and a blocked direct UE--Rx link. For Fig.~\ref{fig:main_results}(a), we set $K=2$, $R_{\rm UE}=2$, and $N=64$; for Fig.~\ref{fig:main_results}(b), we set $N=64$ and the data-phase SNR to $10$~dB; and Fig.~\ref{fig:main_results}(c) evaluates the dominant operation count versus $N$. All algorithms are evaluated over the same channel realizations and with the same joint-detection log-det SE metric. For the simulations, the coefficients of the channels $\ma{G}$ and $\ma{H}$ are generated as independent and identically distributed zero-mean circularly symmetric complex Gaussian random variables, and we assume perfect CSI knowledge.

To provide a strong baseline reference for the proposed method, we consider a multi-start AO benchmark. Specifically, an AO method is run from multiple independent random initializations of the RIS phase vector, and the solution with the largest final joint-detection objective is selected. This strategy mitigates the dependence of AO on the initial RIS configuration. Similar procedures have been used in \cite{guo2020weighted}. In addition, we include a Wu--Liu-inspired \cite{wu2022lowcomplexity} baseline as a representative benchmark, which allows us to assess the performance loss incurred when the RIS design is restricted to a finite set of predefined phase configurations.

The \ac{SE} is evaluated under \ac{JD} as the log-det rate of the  effective channel after transmit precoding, RIS phase shifting, and receive combining. Specifically, for each channel realization, we define $\mathbf{A}=\mathbf{W}^{\text{H}}\mathbf{G}\operatorname{diag}(\mathbf{s})\mathbf{H}\mathbf{Q}$, with $\mathbf{A}\in\mathbb{C}^{R\times R}$. The corresponding sum spectral efficiency is computed as $\mathrm{SE}=\log_2\det\!\left(\mathbf{I}_{R}+\frac{\rho}{R}\mathbf{A}\mathbf{A}^{\text{H}}\right)$, where $\rho$ denotes the transmit SNR in linear scale. The factor $1/R$ accounts for uniform power allocation over the $R$ jointly detected streams. Table \ref{tab:complexity_parameters} describes the specific parameters for each algorithm.

Fig.~\ref{fig:main_results} compares the proposed MS-TAO method with the Wu--Liu-inspired codebook baseline and the multi-start AO benchmark. The results highlight the performance--complexity tradeoff among the considered schemes under the joint-detection setting. In
Fig.~\ref{fig:main_results}(a), we show the sum spectral efficiency as a function of the SNR for fixed $K$ and $R_{\rm UE}$. As expected, all methods exhibit an increasing spectral efficiency as the SNR increases. The proposed MS-TAO method closely follows the multi-start AO benchmark over the entire SNR range, with only a small performance gap. This indicates that the tensor-based alternating updates are able to extract most of the available joint-detection gain without requiring multiple random initializations. In contrast, the Wu--Liu-inspired codebook baseline consistently achieves a lower \ac{SE}. This  is expected since the codebook  restricts the RIS phase vector to a finite set of predefined steering candidates. 

\begin{table}[!t]
\centering
\caption{Algorithm parameters used in the complexity evaluation.}
\label{tab:complexity_parameters}
\footnotesize
\renewcommand{\arraystretch}{1.2}
\setlength{\tabcolsep}{5pt}
\begin{tabular}{l c}
\hline
\textbf{Algorithm} & \textbf{Parameter values} \\
\hline
Proposed MS-TAO &
$I_{\rm TAO}=30$
\\
\hline
Wu--Liu-inspired codebook design &
$N_c=N,\quad I_{\rm WQ}=10$
\\
\hline
Multi-start AO benchmark &
$N_{\rm st}=20,\quad I_{\rm BL}=30,\quad I_{\rm RIS}=25$
\\
\hline
\end{tabular}
\end{table}

In Fig.~\ref{fig:main_results}(b), the \ac{SE} increases with $K$ and $R_{\rm UE}$ in the considered antenna configuration, indicating that the receiver can exploit the additional jointly detected spatial streams. For both values of $R_{\rm UE}$, the proposed MS-TAO method remains close to the multi-start AO benchmark and consistently outperforms the Wu--Liu-inspired baseline \cite{wu2022lowcomplexity}. The performance gap of the codebook method is mainly due to its non-adaptive RIS selection mechanism: the RIS phase vector is chosen from a predefined grid, whereas the proposed MS-TAO directly updates the RIS phase vector globally by exploiting the channel tensor structure.
Fig.~\ref{fig:main_results}(c) compares the dominant computational complexity as a function of the number of RIS elements. This result highlights the main advantage of the proposed MS-TAO method. While the multi-start AO benchmark and the Wu--Liu-inspired baseline \cite{wu2022lowcomplexity} become increasingly expensive as the RIS size grows, the proposed MS-TAO method exhibits substantially more favorable scaling, namely linear scaling with $N$, whereas the multi-start AO benchmark scales quadratically with the RIS size. The Wu--Liu-inspired baseline scales as $N_{\rm c}N^2$, which becomes cubic in $N$ under the setting $N_{\rm c}=N$ used in Table~\ref{tab:complexity_parameters}.


In summary, the proposed MS-TAO method offers the best performance--complexity trade-off among the considered schemes. Its performance remains very close to that of the multi-start AO benchmark while requiring a significantly lower computational complexity. This renders the proposed method particularly attractive for large-RIS multi-user multi-stream systems, where computational efficiency, scalability with the number of RIS elements, and reduced dependence on repeated random initializations are critical design requirements.

\vspace{-0.5cm}
\section{Conclusion}
This letter presents a tensor-based beamforming framework for uplink RIS-assisted multi-user multi-stream \ac{MIMO} systems. Exploiting the multilinear structure of the combined channel tensor, the proposed MS-TAO jointly designs the receive combiner, the block-diagonal users' precoders, and the RIS phase vector. Numerical results showed that MS-TAO attains \ac{SE} close to multi-start AO benchmarks with substantially lower computational cost and consistently outperforms the Wu--Liu-inspired codebook baseline \cite{wu2022lowcomplexity}. This shows the benefit of optimizing the beamformers directly by exploiting the composite tensor structure, especially in large-RIS settings.

\vspace{-0.5cm}

\bibliographystyle{IEEEtran}
\bibliography{ref_adjusted}

@inproceedings{sokal2025tensorRIS,
  author    = {B. Sokal and A. L. F. de Almeida and M. Haardt},
  title     = {Joint Active and Passive Beamforming Design for Multi-Stream {MIMO} {RIS} Systems: A Tensor-Based Approach},
  booktitle = {Proc. IEEE Int. Workshop Comput. Adv. Multi-Sensor Adapt. Process. (CAMSAP)},
  address   = {Punta Cana, Dominican Republic},
  month     = dec,
  year      = {2025},
}

@article{telatar1999capacity,
  title={Capacity of multi-antenna {G}aussian channels},
  author={Telatar, Emre},
  journal={Eur. Trans. Telecommun.},
  volume={10},
  number={6},
  pages={585--595},
  year={1999}
}

@article{verdu2002spectral,
  title={Spectral efficiency in the wideband regime},
  author={Verdú, Sergio},
  journal={IEEE Trans. Inf. Theory},
  volume={48},
  number={6},
  pages={1319--1343},
  year={2002}
}

@ARTICLE{Gil2021,
  author={G. T. {de Araújo} and A. L. F. {de Almeida} and R. {Boyer}},
  journal={IEEE J. Sel. Topics Signal Process.},
  title={Channel Estimation for Intelligent Reflecting Surface Assisted {MIMO} Systems: A Tensor Modeling Approach},
  year={2021},
  volume={15},
  number={3},
  pages={789--802},
  month={Apr.},
  doi={10.1109/JSTSP.2021.3061274}
}

@ARTICLE{MA2021,
  author={X. {Ma} and S. {Guo} and H. {Zhang} and Y. {Fang} and D. {Yuan}},
  journal={IEEE Trans. Wireless Commun.},
  title={Joint Beamforming and Reflecting Design in Reconfigurable Intelligent Surface-Aided Multi-User Communication Systems},
  year={2021},
  volume={20},
  number={5},
  pages={3269--3283},
  month={May},
  doi={10.1109/TWC.2020.3048780}
}

@ARTICLE{WeiLi:2021,
  author={Wei, Li and Huang, Chongwen and Alexandropoulos, George C. and Yuen, Chau and Zhang, Zhaoyang and Debbah, Mérouane},
  journal={IEEE Trans. Commun.},
  title={Channel Estimation for {RIS}-Empowered Multi-User {MISO} Wireless Communications},
  year={2021},
  volume={69},
  number={6},
  pages={4144--4157},
  month={Jun.},
  doi={10.1109/TCOMM.2021.3063236}
}

@article{PARAFAC,
  title={Foundations of the PARAFAC procedure: Models and conditions for an" explanatory" multimodal factor analysis},
  author={Harshman, Richard A and others},
  year={1970},
  journal={UCLA working papes in Phonetics},
  volume={16},
  pages={1-84}
}

@article{you2020energy,
  title={Energy efficiency and spectral efficiency trade off in {RIS}-aided multiuser {MIMO} uplink transmission},
  author={You, Li and Xiong, Jiayuan and Ng, Derrick Wing Kwan and Yuen, Chau and Wang, Wenjin and Gao, Xiqi},
  journal={IEEE Trans. Signal Process.},
  volume={69},
  pages={1407--1421},
  year={2020},
  month={Dec.},
  publisher={IEEE}
}

@article{ardah2021trice,
  title={{TRICE}: A channel estimation framework for {RIS}-aided millimeter-wave {MIMO} systems},
  author={Ardah, Khaled and Gherekhloo, Sepideh and de Almeida, Andr{\'e} L. F. and Haardt, Martin},
  journal={IEEE Signal Process. Lett.},
  volume={28},
  pages={513--517},
  year={2021},
  month={Feb.},
  publisher={IEEE}
}

@ARTICLE{gil2022,
  author={de Ara{\'u}jo, Gilderlan T. and Gomes, Paulo R. B. and de Almeida, Andr{\'e} L. F. and Fodor, G{\'a}bor and Makki, Behrooz},
  journal={IEEE Wireless Commun. Lett.},
  title={Semi-Blind Joint Channel and Symbol Estimation in {IRS}-Assisted Multi-User {MIMO} Networks},
  year={2022},
  volume={11},
  number={7},
  pages={1553--1557},
  month={Jul.}
}

@book{tse2005fundamentals,
  title        = {Fundamentals of Wireless Communication},
  author       = {Tse, David and Viswanath, Pramod},
  publisher    = {Cambridge University Press},
  year         = {2005}
}

@article{wu2019intelligent,
  title={Intelligent reflecting surface enhanced wireless network via joint active and passive beamforming},
  author={Wu, Qingqing and Zhang, Rui},
  journal={IEEE Trans. Wireless Commun.},
  volume={18},
  number={11},
  pages={5394--5409},
  year={2019}
}

@article{huang2019reconfigurable,
  title={Reconfigurable intelligent surfaces for energy efficiency in wireless communication},
  author={Huang, Cunhua and Zappone, Alessio and Alexandropoulos, George C and Debbah, M{\'e}rouane and Yuen, Chau},
  journal={IEEE Trans. Wireless Commun.},
  volume={18},
  number={8},
  pages={4157--4170},
  year={2019}
}

@article{bjornson2020reconfigurable,
  title={Reconfigurable intelligent surfaces: A signal processing perspective with wireless applications},
  author={Bj{\"o}rnson, Emil and Sanguinetti, Luca},
  journal={IEEE Signal Process. Mag.},
  volume={37},
  number={6},
  pages={135--158},
  year={2020}
}

@inproceedings{wu2022lowcomplexity,
  author    = {Wu, Yu-Tse and Liu, Kuang-Hao},
  title     = {{Low-complexity Beamforming Design for RIS Communications over Correlated Channels}},
  booktitle = {Proc. IEEE Wireless Commun. Netw. Conf. (WCNC)},
  pages     = {1069--1074},
  year      = {2022},
  doi       = {10.1109/WCNC51071.2022.9771861}
}

@article{guo2020weighted,
  author  = {Guo, Huayan and Liang, Ying-Chang and Chen, Jie and Larsson, Erik G.},
  title   = {{Weighted sum-rate maximization for reconfigurable intelligent surface aided wireless networks}},
  journal = {IEEE Trans. Wireless Commun.},
  volume  = {19},
  number  = {5},
  pages   = {3064--3076},
  month   = may,
  year    = {2020},
  doi     = {10.1109/TWC.2020.2970061}
}

\end{document}